\documentclass[preprint,aps, preprintnumbers, nofootinbib]{revtex4}
\usepackage[dvips]{graphics}

\newcommand{\gsim}[2]{
\setlength{\unitlength}{12pt}
\begin{picture}(1.4,1.)
\put(.7,-0.3){\makebox(0.0,1.)[t]{$>$}}
\put(.7,-0.3){\makebox(0.0,1.)[b]{$\sim$}}
\end{picture}#2}

\begin{document}

\title{Entanglement in holographic dark energy models}

\author{R. Horvat}
\email{horvat@lei3.irb.hr}
\address{Rudjer Bo\v{s}kovi\'{c} Institute, P.O.B. 180, 10002 Zagreb,
Croatia}

\begin{abstract}
We study a process of equilibration of holographic dark energy (HDE) with
the cosmic horizon around the dark-energy dominated epoch. This process 
is characterized by a huge amount of
information conveyed across the horizon, filling thereby a large gap in
entropy between the system on the brink of 
experiencing a sudden collapse to a black hole
and the black hole itself. At the same time, even in the absence of
interaction between dark matter and dark energy, such a process marks a
strong jump in the entanglement entropy, measuring the quantum-mechanical
correlations between the horizon and its interior. Although the
effective quantum field theory (QFT) 
with a peculiar relationship between the UV and IR
cutoffs, a framework underlying all HDE models, may formally account for 
such a huge shift in the number of distinct quantum states, we show that 
the scope of such a framework becomes tremendously restricted, devoiding it 
virtually any 
application in other cosmological epochs or particle-physics phenomena. The
problem of negative entropies for the non-phantom stuff is also discussed.    
\end{abstract}

\newpage

\maketitle

A great variety of diverse models have been invoked to shed light on the
present phase of accelerated expansion of the universe \cite{1}. Amongst
them a field-theoretical setup with the encoded information from quantum
gravity, leading to a novel variable cosmological constant (CC) approach and
generically dubbed that of `holographic dark energy' (HDE), has recently
triggered a lot of interest \cite{2, 3, 4}. Besides the dark energy 
problem, such a
framework has also proved to have a potential to  
shed light both on the `old' CC problem
\cite{5} and the `cosmic coincidence problem' \cite{6}.  

A field-theoretical framework \cite{7} underlying all
HDE models 
describes a rather peculiar object. For a region of the size $L$ (providing an
IR cutoff) it describes a system on the brink of experiencing a sudden
collapse to a black hole in that its energy density is the same as that of
the black hole of the same size. As opposed to that, however, its entropy is
tremendously less than the corresponding black hole entropy. If we are to
describe such a system using an ordinary quantum field theory (QFT), and 
since black holes appear to involve a vast number of states not describable
within it \cite{8}, then the QFT entropy $\sim  $ $L^3 {\Lambda }^3 $ should 
obey at saturation \cite{7}
\begin{equation}
L^3 {\Lambda }^3 \sim L^{3/2} M_{Pl}^{3/2} \sim (S_{BH})^{3/4} \ll S_{BH}
\;,
\end{equation}
where $\Lambda$ is the UV cutoff and  
$S_{BH}$ is the entropy of a black hole of the size $L$, $S_{BH} \sim L^2
M_{Pl}^2$. This creates thus a huge gap in entropy between the two systems 
having the same size and energy. Since the entropy in QFT scales 
extensively, it is clear that in a cosmological setting 
$\Lambda $ should be promoted to a varying
quantity (some function of $L$ to manifest the UV/IR connection),
in order (1) not to  be violated during the course of the
expansion. This gives the maximum energy density in the
effective theory, $\rho_{\Lambda } \sim {\Lambda }^4 $, to be $\rho_{\Lambda
} \sim  L^{-2}M_{Pl}^2 $. Obviously, $\rho_{\Lambda }$ is the energy density
corresponding to a zero-point energy and the cutoff $\Lambda $. The main
reason of why the above HDE model is so appealing in possible
description
of dark energy is  that $\rho_{\Lambda }$ as given above gives
the right amount of  dark energy in the
universe at present, provided $L \simeq H^{-1}$, where $H$ is the Hubble
parameter. This also eliminates the need for fine-tuning in the `old' CC
problem \cite{7}.

Another consequence of placing the system described by (1) in a
cosmological setting is the existence of event horizons induced by the
energy density $\rho_{\Lambda} \sim L^{-2}M_{Pl}^2$. This means that our 
isolated
system actually splits into two subsystems: the cosmological horizon and the
stuff inside the horizon including the system described by (1). Obviously,
the entropy (1) of the isolated system has nothing to do with 
thermodynamical entropies; it rather represents the 
so-called fine-grained entropy, 
which stays exactly conserved in whatever setting the system is placed in.
With the two subsystems, however, the horizon and the interior  are expected 
to become 
entangled, and also thermalized  
sooner or 
later, leading to non-zero thermodynamical entropies as well as to the growth
of the entanglement entropy. Thermodynamical entropies are thus additive 
but not conserved (owing to the generalized  second law of 
gravitational thermodynamics), whereas the fine-grained (or entanglement)
entropies are conserved but not additive (the fine-grained entropy of the
whole system always stays at the value (1)). 

To get a feeling of why a
process of equilibration of the system (1) with the cosmological horizon is
so abrupt and violent, note that an ordinary QFT is capable of describing a
system at a temperature $T$ provided 
\begin{equation}
\Lambda \sim T >> L^{-1} \;.
\end{equation}
On the other hand, the instantaneous horizon temperature is $T_{hor} \sim
L^{-1}$, which irrespective of the choice for the IR cutoff should at
present be $\sim 10^{-33} $ eV. From the relationship between the cutoffs
\cite{7} 
\begin{equation}
\Lambda^4 \simeq L^{-2}M_{Pl}^2 \;,
\end{equation}        
one however gets $\Lambda \sim 10^{-3}$ eV. It is just this huge
disparity in temperatures between the dark energy stuff and the cosmic
horizon that makes this process so peculiar. In the equilibration process 
the UV and the IR cutoff must thus come very close together,
$\Lambda \sim L^{-1}$, which does have
severe consequences for the underlying QFT.    

The present paper is about ${\sl internal}$ ${\sl
inconsistencies}$ (as sketched above) inherent to ${\sl any}$
HDE model (independent of the choice for the IR cutoff $L^{-1}$) and underlied
by the original theoretical framework \cite{7}. The problem emerges when
a non-black hole object like the HDE [having the same energy as the black
hole
for a given size but tremendously less
entropy as given by (1)] is placed in a cosmological setting. Besides
causing the universe to accelerate at present times, such a placing does
inevitably trigger the formation of a cosmological horizon - a cosmological
black hole. The root of the
problem is how to thermalize 
an inherently non-black hole object (HDE) with
the cosmic horizon (a black hole object whose entropy measures our ignorance
of what lies beyond). The paper explores the various aspects of the
above problem, showing, most importantly, that quantum correlations between the 
horizon and the interior (consisting mostly of HDE in the dark-energy 
dominated epoch) and embodied in entanglement entropy turn out to be hopelessly
tiny in order to trigger the thermalization process.

Below we first study the process of how the HDE gets thermalized 
with the cosmic horizon near the dark-energy dominated epoch purely on
phenomenological grounds. Then we propose
how the underlying QFT should be changed in order to account for such a 
violent process. Finally, we stress the restrictive scope of such a QFT. In
order to avoid any influence of other components on the thermalization
process, we shall expose our ideas with the aid of the non-interacting Li's
model \cite{3}. 

Some thermodynamical aspects of HDE models (the first and the generalized
second law) have already been studied \cite{9, 10, 11, 12, 13, 14}. Usually the
fluid temperature is taken to be at or proportional to the horizon 
temperature. Let us first set up when it is appropriate to choose so. 

Here we state that thermodynamic equilibrium of the HDE with the
horizon gets established if
\begin{equation}
\left |\frac{d_E}{\dot{d}_E} \right| \gsim \; \frac{d_E}{c_{\gamma}} \;,
\end{equation}
where the future event horizon $d_E$ is given by
\begin{equation}
d_{E} = a \int_{a}^{\infty } \frac{da}{a^2 H} \;,
\end{equation}
with $a$ being a scale factor. That is, departures from de Sitter space
should be small enough so that the RHE of (4) is always larger than the 
light-crossing
time of the radius $d_E$. Thermodynamic equilibrium having once been
established at such time, it continues to exist provided the heat capacity
for the whole system is positive-definite \cite{9, 15}. Since the heat
capacity of the horizon is negative, the heat
capacity of the dark energy fluid should be positive (and larger in absolute
value).

Let us now see how the above postulates 
work for the popular Li's model \cite{3}. In a
two-component universe $\rho_{\Lambda}$ evolution is governed
by \cite{3, 15}
\begin{equation}
\Omega_{\Lambda }^{'} = \Omega_{\Lambda }^2 ( 1 - \Omega_{\Lambda }) \left
[\frac{1}{\Omega_{\Lambda }} + \frac{2}{c \sqrt{\Omega_{\Lambda }}} \right ]
\;,
\end{equation}
where the prime denotes the derivative with respect to $lna $. In (6)
$\Omega_{\Lambda } = \rho_{\Lambda }/\rho_{crit} $, where $\rho_{crit}$ is
the
critical density and $\rho_{\Lambda } $ was parametrized as $\rho_{\Lambda }
=(3/8 \pi ) c^2 M_{Pl}^2 L^{-2}$, with 
a constant parameter $c$ of order one and $L = d_E$. Also $\Omega_{\Lambda }
+ \Omega_X = 1$, with $X$ being either matter or radiation. Combining (6)
with (4) for the matter case one arrives at
\begin{equation}
\left |\frac{\sqrt{\Omega_{\Lambda }}}{c -\sqrt{ \Omega_{\Lambda }}} \right| 
\gsim \; 1 \;.  
\end{equation}
Employing $c=1$\footnote{Although a restriction on $c^2$ under the combined
phenomenological constraints obtained recently \cite{17} slightly favor
phantom behavior ($c^2 < 1$), our entropic arguments favor $c=1$ (see
below). Anyhow $c=1$ taken in (7) is for the illustration purposes
only.}, one obtains $\Omega_{\Lambda } > 1/4$. This is what is to be
expected: the HDE enters the thermodynamic equilibrium with  the horizon 
somewhere around the onset of the dark-energy dominated epoch. To see that
this is by no means so for earlier cosmological epochs, we note that (6) is
also capable of describing epochs where $\rho_{\Lambda }$ occupies only a
tiny fraction of the total energy density. In particular, in that limit 
$\Omega_{\Lambda}  <<1$,
the solution of (6) in the radiation-dominated universe reads
\begin{equation}
L(a) \simeq M_{Pl} \; \rho_{rad 0}^{-1/2} \; a^{3/2} \;,
\end{equation}  
where $\rho_{rad 0}$  denotes the radiation energy density at the present
time. Using (8) one can be easily convinced that (4) is far from being
satisfied wherever in the radiation-dominated epoch of the universe.

Hence if we 
trust the qualitative criterion (4), then the HDE becomes thermalized
with the horizon near the onset of the dark-energy dominated epoch. This means
equalizing of the temperatures, that is, rapprochement of the cutoffs,
$\Lambda$ and $L^{-1}$. But it is obvious right away that the theoretical
setup as given by (1) and (3) is not capable to support this scenario. By 
setting $\Lambda \sim
L^{-1}$ in (3), one gets $\Lambda \sim L^{-1} \sim M_{Pl}$, and we need this
not in the Planck-time era but some $10^{60}$ Planck times later.

There is also a more physical argument against (1) and (3): the entanglement
entropy. The entanglement entropy in the present context would measure
quantum-mechanical correlations between the horizon and its interior. As
soon as the physical horizon forms, and consequently an interior observer
lacks any information about the space outside the horizon, both the horizon
entropy (the black hole entropy) and the entanglement entropy become
nonzero. When the overall state is pure or near-pure, the entanglement
entropy should behave nonextensively, that is, should depend only on the
surface of the horizon separating the interior from the rest. On the other
hand, quantum correlations between the subsystems in any local QFT are taken
care of by the UV cutoff. Consequently, we have
\begin{equation}
S_{ent} \sim \Lambda^2 L^2 \;.
\end{equation} 
Using (3) one arrives at
\begin{equation}
S_{ent} \sim \ L M_{Pl} \;.    
\end{equation}
Comparing (10) with other two type of entropies involved in the problem,
$S_{HDE} \sim L^{3/2} M_{Pl}^{3/2}$ and that of black holes $S_{BH} \sim L^2
M_{Pl}^2$, $S_{ent}$ is by far the least
one near the present epoch, leading to the prominent hierarchy 
\begin{equation}
L M_{Pl} << L^{3/2} M_{Pl}^{3/2} << L^2 M_{Pl}^2 \;.
\end{equation}
The physical interpretation of (11) is pretty obvious. The two subsystems do
interact extremely weakly so that the thermalization process cannot be
initiated. Obviously, the thermodynamics of the HDE models near the present
epoch is not possible within the original theoretical framework \cite{7}. 

Now we can ask: is it possible to change the original QFT framework to 
account for a huge transfer of entropy across the horizon, needed to start a
thermalization process? One possibility \cite{18}, staying purely within the
realm of effective QFT, is to develop the original theory with a large number
of particle species, $N >>1$. The basic idea is that with $N >>1$ the energy
density $\rho_{\Lambda} \sim  L^{-2}M_{Pl}^2$ stays intact, whereas both
entropies, $S_{ent}$ and $S_{HDE}$, now become some increasing functions of
the number of field species. When the  maximal allowable limit for $N$ is
approached, both $S_{ent}$ and $S_{HDE}$ begin to sustain the black
hole entropy, $L^2 M_{Pl}^2$. The huge gap in entropies between the HDE
object and black holes is thus being populated when $N$ is increased. With $N
>>1$ (3) now gets modified to
\begin{equation}
N \Lambda^4 \simeq L^{-2}M_{Pl}^2 \;.
\end{equation}

Using a criterion that thermal equilibrium between the HDE and the horizon
gets established when 
\begin{equation}
S_{ent} \sim S_{BH} \;,
\end{equation}
one can determine $S_{ent}$ using (12), and than find $N$ from (13). Noting
that $S_{ent}$ also scales with $N$, one obtains\footnote{Indeed, by
plugging (14) back into (12), a wanted result $\Lambda \sim L^{-1}$ is
obtained.}
\begin{equation}
N \sim L^2 M_{Pl}^2 \;,
\end{equation}    
a really huge number ($\sim 10^{122}$) if $L$ is taken of order of the 
horizon distance at
present. Within the same framework that much large $N$ would cause 
problems with 
overproduction of gravitinos in a low-entropy post reheating epoch
\cite{18}. Also, a bound $N_{max} \simeq 10^{32}$ was obtained in alike
theories by noting that we have not seen any strong gravity in the particle
collisions \cite{19} \footnote{The QFT of KK particles
(or equivalently for four-dimensional
models with large-$N$ species as covered in \cite{19}) only takes cares of
the UV
cutoff - how it gets reduced in the presence of a large number of particle
species. It provides the benchmark value $(N  \simeq 10^{32})$, obtained by
noting
that we have not seen any strong gravity in the particle collisions.
On the other hand the present QFT setup deals both with the UV and
IR cutoffs, furthermore the scenario does exhibit a peculiar sort of UV/IR
mixing, meaning that this way an information from quantum gravity
becomes encoded in such
a QFT framework (the main motivation for such a modification of the
effective QFT framework being of course the compliance of the QFT with the
holographic
bounds). The two large-$N$ approaches coincide only if the
cutoffs coincide, i.e., if $\Lambda \simeq L^{-1}$, which is nothing but the
black-hole limit (this is why the authors of [19] indicated a
non-perturbative nature of their bound). It can be shown that the
fine-grained entropy as given by (1) becomes $N$-dependent and begins to
sustain the black hole entropy for the maximal $N$, which in the present
scenario amounts $\Lambda \simeq L^{-1}$.}. Even better limits can be inferred in the present
framework when considering some particle-physics phenomena \cite{20}. 
In addition, with $N$ as a running number
as given by (14), a question regarding internal consistency of the large-$N$ 
framework
does also arise. Namely, in the Li's model $d_E = L \sim a^{1 - 1/c}$ 
when dark energy
dominates, and the horizon area is non-decreasing with time 
for the non-phantom stuff $(c \geq 1)$. This means that 
in order to maintain thermal equilibrium $N$ from (14) should
grow without limit as time goes by if $c > 1$, jeopardizing thereby 
the internal consistency of
the framework. In order to keep the internal consistency one thus has to
resort to the de Sitter limit in the infinitely far future, i.e., $c =1$. 
In this 
case $N$ saturates asymptotically to a finite
number. It is interesting to see how the
internal consistency of the large-$N$
HDE framework singles out $c =1$ when applied to the Li's model.

If one still insists that the large-$N$ HDE framework (or whatever other
unknown mechanism) is capable to bring a fluid with $\rho_{\Lambda}
\sim  L^{-2}M_{Pl}^2$ in thermal equilibrium with the horizon 
near the present epoch, then one can speak for the first time of 
thermal (or coarse grained) entropies. They can be
determined with the aid of the first law of thermodynamics
\begin{equation}
T_{hor}dS_{\Lambda } = d(\rho_{\Lambda }V) + p_{\Lambda }dV \;,
\end{equation}
where $T_{hor} = 1/(2\pi L)$ is the horizon 
temperature, $V= (4\pi /3) L^3$ and
$p_{\Lambda } = w_{\Lambda } \rho_{\Lambda }$. One obtains
\begin{equation}
dS_{\Lambda } = \pi M_{Pl}^2 c^2 (1 + 3w_{\Lambda }) L dL \;.
\end{equation}
Noting that $w_{\Lambda } = -1/3 -2\sqrt{\Omega_{\Lambda}}/3c$, it can be
seen   that with the integration of (16) one necessarily  deals with
negative entropies for the non-phantom stuff ($c >1$). In this case the
horizon area grows without limit towards future and the constant of
integration cannot be chosen as to make up for this negative contribution.
Thus a non-phantom fluid effectively 
behaves as a  phantom-fluid whose entropy is 
always negative \cite{21}.
We see again that a true CC limit in the infinitely far future  
($c = 1$) is singled out. In this limit 
the horizon
area approaches asymptotically a constant value, so the constant of
integration can be appropriately chosen as to make the  total contribution
positive. 

Finally, a note on maintaining the thermal equilibrium. The heat capacity 
is defined as $C_{X} = T(\partial S_{X}/\partial T)$, with $X$ either the 
horizon or the HDE stuff. Since $C_{hor} = -2\pi M_{Pl}^2 L^2$ and
$C_{\Lambda}
= 2\pi c \sqrt{\Omega_{\Lambda}} M_{Pl}^2 L^2$, one sees that the
requirement that the sum be positive boils down to $c
\sqrt{\Omega_{\Lambda}} >1$. This shows  that only 
the case away from the true CC
limit ($c > 1$) is relevant for maintaining thermal equilibrium.        

In conclusion, we have found out  that the effective QFT framework, underlying
all HDE models, is not capable to describe the
holographic dark energy component in thermal equilibrium with the cosmic
event horizon around the present time, a process which can be successfully
described phenomenologically. 
The dark energy component inside the
horizon would have to have the enormously larger  entropy as well as 
the enormously
smaller temperature than what is consistent with the underlying theoretical 
setup.
When the framework is developed with a large number of particle species, the
dark energy entropy tends to increase with $N$ while the UV cutoff (a measure
of the temperature)  tends to decrease with $N$. The present-day
cosmological requirement on $N$ is however so huge to be consistent with 
other 
phenomenological  constraints on the number of particle species. In addition,     
the UV cutoff
is so hugely diminished that such a QFT is not capable to describe virtually 
any
relevant physics. For instance, such a theory is not capable to describe even
thermal photons of the universe at present, at a temperature $\sim
10^{-4}$ eV. If one is contented with the phenomenological description of
the process only, then this would entail a problem where negative entropies
do arise for the non-phantom component. The origin of this problem lies in the
fact that the energy density of the HDE is devoid of a true constant term
\cite{22}. Unfortunately, holography always does away with such a
constant term in the energy density. In contrast, renormalization-group running cosmologies
\cite{23}, besides having the same variable part of the energy density as 
the HDE component, are always accompanied with such a term. The constant
serves to prevent the horizon area to grow with time without limit. For the
present case this is only the case for a singular point ($c =1$) in the
parameter space. For another argument supporting c=1 in the far future, see
\cite{24}. On the other hand, maintaining thermal 
equilibrium with the horizon is
not consistent with this point. The internal inconsistencies in HDE models
found here adds to the previous ones having been discussed earlier
\cite{25}.

{\bf Acknowledgment. } This work was supported by the Ministry of Science,
Education and Sport
of the Republic of Croatia under contract No. 098-0982887-2872.

\end{document}